# Covalent bonding and magnetism in cuprates

Andrew C. Walters[1,2,3], Toby G. Perring[1,2], Jean-Sébastien Caux[4], Andrei T. Savici[5], Genda D. Gu[5], Chi-Cheng Lee[5], Wei Ku[5], and Igor A. Zaliznyak[5]

[1] *ISIS Facility, Rutherford Appleton Laboratory, Chilton, Didcot OX11 0QX, UK*

[2] *Department of Physics, University College London, Gower Street, London WC1E 6BT*

[3] *London Centre for Nanotechnology, 17-19 Gordon Street, London WC1H 0AJ, UK*

[4] *Institute for Theoretical Physics, University of Amsterdam, 1018 XE Amsterdam, The Netherlands*

[5] *CMP&MS Department, Brookhaven National Laboratory, Upton, New York 11973, USA*

**The importance of covalent bonding for the magnetism of 3d metal complexes was first noted by Pauling[1] in 1931. His point became moot, however, with the success of the ionic picture of Van Vleck[2], where ligands influence magnetic electrons of 3d ions mainly through electrostatic fields. Anderson's theory of spin superexchange[3] later established that covalency is at the heart of cooperative magnetism in insulators, but its energy scale was believed to be small compared to other inter-ionic interactions and therefore it was considered a small perturbation of the ionic picture[4]. This assertion fails dramatically in copper oxides[5], which came to prominence following the discovery of high critical temperature superconductors (HTSC). Magnetic interactions in cuprates are remarkably strong and are often considered the origin of the unusually high superconducting transition temperature, Tc. Here we report a detailed survey of magnetic excitations in the one-dimensional cuprate $Sr_2CuO_3$ using inelastic neutron scattering (INS). We show that although the experimental dynamical spin structure factor is well**

**described by the model S=1/2 nearest-neighbour Heisenberg Hamiltonian typically used for cuprates, the magnetic intensity is modified dramatically by strong hybridization of Cu 3d states with O p states, showing that the ionic picture of localized 3d Heisenberg spin magnetism is grossly inadequate. Our findings provide a natural explanation for the puzzle of the missing INS magnetic intensity in cuprates and have profound implications for understanding current and future experimental data on these materials.**

Over the last twenty years, the magnetic properties of cuprates have been studied extensively by theorists and experimentalists alike. These systems are usually described within the antiferromagnetic Mott insulator model, in which the unpaired electrons are localized on the $Cu^{2+}$ ions because of the overwhelming cost in the on-site Coulomb interaction energy, $U$, associated with the charge transfer between the Cu sites, a strong correlation phenomenon. Virtual electron hopping, which in the one-band Hubbard model of a Mott insulator often adopted for cuprates[6] is quantified by the transition matrix element, $t$, results in antiferromagnetic exchange. For $t << U$, electron spins form the only low-energy electronic degrees of freedom. Their properties are well approximated by the spin-1/2 Heisenberg Hamiltonian on the lattice[4],
$H = J \sum_{i(nn)j} \mathbf{S}_i \mathbf{S}_j$, with the nearest-neighbour exchange coupling $J \approx 4t^2/U$.

This description conveniently splits the problem of electronic magnetism in the Mott insulator into two parts[4]. The first deals with electron transfer between the neighbouring sites of the crystal lattice, which is determined by the overlap integral ($\sim t$) of the wave functions occupied by the unpaired electrons and leads to the Hubbard model or the Heisenberg spin Hamiltonian. The second concerns the form of the electronic Wannier wave functions, that is, the shape of the spin magnetization cloud associated with each electron in the obtained Hubbard or Heisenberg model. It is determined by the crystal electric field and the hybridization of the unfilled 3d orbitals with those of the surrounding p-orbitals of the anions – the covalency – and is usually addressed by first principles calculations. The first part provides a description of the cooperative behaviour of electronic spins – spin order and spin excitations, while the second relates them to the behaviour of the magnetization density in the crystal, which is measured in actual experiments.

In most cases the hybridization effects yield only a small, 10% to 20% covalency correction and were traditionally considered unimportant for magnetism in Mott insulators[4]. They were nevertheless noticed in the precise neutron diffraction study of NiO by Alperin[7] and were elegantly explained by Hubbard and Marshall[8]. Modern polarized neutron diffraction experiments in systems with strongly correlated electrons also often find small magnetic moment density on the ligand site and the corresponding ~ 10% covalent reduction of the magnetic moment at the 3d site[9].

That covalency plays a fundamental role and should not be so easily dismissed in the cuprates became quite clear when the state of the doped holes, leading to the HTSC, was considered. Zhang and Rice have shown[5] that strong Cu(3d)-O(2p) hybridization in fact defies Hund's rule and leads to an unusual singlet state of the doped hole, whose wave function is mainly localized on oxygens, instead of the S=1 $(3d)^8$ $Cu^{3+}$ state expected in the ionic picture. The extraordinary strength of Cu-O covalent bonding is also responsible for the record-high antiferromagnetic exchange coupling found in the undoped parent cuprates, $J \approx 1500$ K[10,11] in the two-dimensional (2D) $La_2CuO_4$ and $J \approx 2600$ K[12] in its 1D chain relative $SrCuO_2$. Such a distinctively dominant magnetic energy scale immediately suggests that it is a key player in the mechanisms of the HTSC in cuprates. The precise nature of this role – whether magnetic excitations provide the interaction that pairs doped charges, or dynamical spin correlations enable the effective fractionalization of the doped charges and negate their Fermion nature at low energy, or that they cause phase segregation favouring superconductivity – is a matter of ongoing debate[13,14]. The close relationship between the magnetism and HTSC is supported by a body of INS studies, in particular by discoveries of the sharp resonance peak of magnetic excitations and the low-energy incommensurate scattering whose temperature, doping, and magnetic field dependencies are all correlated with superconductivity[13,14,15,16,17].

Neutrons are an ideal probe of microscopic magnetism in condensed matter, as they interact with electronic magnetic moments directly, via the dipole-dipole force. The neutron magnetic scattering cross-section[18] can be conveniently factored into the product of the dynamical spin correlation function, $S(\mathbf{Q},E)$, which is determined by the cooperative behaviour of electronic spins described by the Hubbard, $t$-$J$ or Heisenberg model, and the square of the magnetic form factor, $|F(\mathbf{Q})|^2$. $F(\mathbf{Q})$ is the Fourier transform of the electronic magnetization cloud associated with each spin, that is, its Wannier wave function, and includes the covalency effects. According to LDA calculations[19,20,21] and the general expectations outlined above, their impact on magnetic INS intensity should be extremely strong in cuprates.

Quantitative analysis of the measured magnetic INS intensities and their changes upon entering the superconducting state can be used to evaluate the corresponding changes in the electronic kinetic and spin-interaction energies. This was deployed in a number of influential studies with the aim of quantitatively testing certain predictions of the Hubbard or the $t$-$J$ model, often used for the cuprates[22,23,24]. The accuracy of such an approach, however, relies entirely on either neglecting the covalency effects on measured INS intensities, or the precise knowledge and account of these effects, which again raises the question of the importance of covalency in cuprates. Surprisingly, neutron diffraction found very little, if any covalent reduction of the ordered moment in cuprates[25,26]. This controversy was pointed out by Kaplan, et. al.[27], but the point was declared moot in view of the perhaps fortuitous agreement between the experiment and the semiclassical theory. Diffraction, however, only probes $|F(\mathbf{Q})|^2$ at a select number of points where magnetic Bragg peaks are present and therefore its sensitivity to the covalency is very limited and constrained by the symmetry of the magnetic structure.

Recent advances in neutron scattering technology, including the development of time-of-flight (TOF) neutron spectrometers such as MAPS at the ISIS spallation source at the Rutherford Laboratory in UK, which was used in our study, have enabled magnetic excitations of energy up to ~ 1 eV to be measured routinely and with great precision. Previous measurements of the magnon dispersion in the 2D cuprate $La_2CuO_4$ were sufficiently sensitive to observe and quantify subtle deviations from the Heisenberg model, and revealed the existence of ring exchange[11]. The analysis of the spectral weight of the spin correlation function based on the measured absolute magnetic intensity, however, has not been as successful. In fact, such analysis is routinely performed using the ionic $Cu^{2+}$ magnetic form factor and thus totally neglects covalency effects. A number of studies [10,11, 12] have consequently found significant deviations of the INS spectral weight from sum rules, which in particular state that the integral of $S(Q,E)$, over $Q$ and energy should equal to $NS(S+1)$, where $N$ is the number of magnetic sites in the sample and $S$ is the spin at each site (S = ½ for cuprates). While covalency is a leading suspect for these outstanding discrepancies, their unambiguous association with the covalent magnetic form factor in planar cuprates is hindered by the absence of a precise theory describing the dynamical behaviour of the 2D spin system. Consequently, other explanations for the missing intensity have been developed, in particular, ascribing it to the salient features of $S(Q,E)$ in the 2D Hubbard model[28].

The situation is very different in 1D, where recent progress in the theory of integrable models has led to the development of an extremely accurate quantitative theory for the dynamical spin correlations[29]. Here we report detailed INS measurement of magnetic excitations in the 1D prototype cuprate material $Sr_2CuO_3$, whose crystal structure is illustrated in Figure 1(a). It features chains of corner-sharing $CuO_4$ plaquettes, which are the building block of 2D HTSC cuprate planes and related spin-ladders, running along the crystallographic $b$-axis. While the direct determination of the in-chain spin exchange coupling is only possible by INS and is reported here, the estimate of $J \approx 250$ meV obtained from optical absorption measurements[30] indicates similar, or stronger degree of covalence as in 2D cuprates. The inter-chain orbital overlaps are extremely small in $Sr_2CuO_3$, resulting in negligible inter-chain hopping and spin coupling and rendering this material the record-holder for one-dimensionality: 3D antiferromagnetism in $Sr_2CuO_3$ only appears below $T_N \approx 5$ K[31], which yields interchain exchange a factor $\geq 10^3$ smaller than the in-chain exchange. Magnetic orbitals and corresponding form factors used in our study are shown in Figure 1(b-h).

Colour contour maps in Figure 2 present an overview of our INS data. They show the normalized intensity of magnetic scattering by the high quality single crystal of $Sr_2CuO_3$ at $T \approx 5.5$ K. The measured non-magnetic signal – due to incoherent and multi-phonon scattering, etc. – has in all cases been subtracted. The data covers two 1D Brillouin zones (BZ) and was not symmetrized. Magnetic scattering corresponding to the multi-spinon triplet continuum[29] emanating from $Q_{chain} = \pm 0.5$, where it is most intense, and extending to $E \sim 600$ meV, is clearly seen repeated in both BZ. The continuum of excitations filling large regions of ($Q_{chain}, E$) phase space is clearly visible in the two upper panels. In the lower two panels it is possible to identify both strong spinon scattering emerging from $Q_{chain} = \pm 0.5$ and weaker returning spinon branches around $Q_{chain} = 0, \pm 1$ with a $|\sin(\pi Q)|$-like dispersion extending up to $\approx 350$ meV.

We analyze our data by fitting the $Q_{chain}$-dependent intensity of 1D constant-energy cuts at different energy transfers to the exact dynamical structure factor for 1D spin-½ Heisenberg Hamiltonian including both the two-spinon and four-spinon contributions to $S(Q_{chain}, E)$ [29]. Although higher-order excitations will also form a part of the excitation continuum, the calculated spectral weight arising from these two contributions is about 98% of the total, which is within our experimental error.

We began by fitting the measured intensities to the exact dynamical structure factor $S(Q_{chain}, E)$ [29] and using the anisotropic magnetic form factor of the $Cu^{2+}$ ion, following the procedure used in previous studies[11,12,25,26]. Only two parameters are varied in such fitting: the in-chain exchange coupling $J$, which determines the overall energy scale of magnetic excitations, and the prefactor $A$, which accounts for the possible statistical and systematic errors of our intensity measurements and ideally should be equal to 1. Examples of the resulting fits are shown by solid lines in the left panel of Figure 3, and the obtained parameters $A$ and $J$ are shown in Figure 4. While the overall agreement with the data in Figure 3, (a)-(d) is reasonable, the intensity prefactor $A = 0.32(3)$ is far too small – it appears as if we are only measuring 1/3 of the predicted intensity. This falls well beyond the most conservative estimates of statistical and systematic errors of our experiment, which are given by the error bars and the point scatter in Figure 4, respectively. A similar factor of 2 to 3 missing intensity was reported in the related chain cuprate $SrCuO_2$ – up to now presenting a puzzle[12]. As mentioned earlier, missing intensity was also reported in planar cuprates[10,11], although in the absence of exact theory for $S(Q,E)$ in 2D it is difficult to quantify its significance.

Anticipating covalency as the prime suspect for these outstanding discrepancies, we deployed a relatively recent development in Density Functional Theory – the implementation of the LDA+U functional, which has been successful in describing magnetic insulating phase of cuprates[19,20,21] – and computed the low-energy Wannier function and the corresponding magnetic form factor for $Sr_2CuO_3$. The results are shown in panels (b) and (d) of Figure 1 and are compared in more detail with the ionic $Cu^{2+}$ magnetic form factor in panels (f)-(h). Figure 1 (b) clearly demonstrates strong hybridization between Cu $d$- and O $p$-states and significant magnetic density on the oxygen sites. The resulting covalent magnetic form factor in Figure 1 (d) reflects this behaviour. A larger extent of the wave function in real space naturally leads to a smaller size of the form factor in wave vector space and consequently smaller INS intensity at the same wave vector.

Fits of the data with the same dynamical structure factor but with the newly calculated covalent magnetic form factor of Figure 1(d) are shown in the right panels (e)-(h) of Figure 3. The better agreement of the new fits with the data for $Q_{chain} \leq 0.5$ is clearly visible upon comparison of the left and the right panels of Figure 3. Most importantly, the value of the prefactor $A$ shown in Figure 4(a) now accounts for $\approx 80\%$ of the expected intensity. This is a change by a factor $\approx 2.5$ – a dramatic correction to the ionic picture. The remaining small discrepancy of 20% is most likely accounted by the Debye-Waller (DW) factor, which results from the disorder of the ionic positions in real materials, and superposes a much weaker wave vector-dependent reduction of coherent scattering intensity. Its effect is illustrated in the inset of Figure 4(a). The values of the exchange coupling $J$ refined in both fits practically coincide, $J = 241(11)$ meV, reflecting the fact that excitation dispersion is sensitive to the positions of peaks in the measured intensities in ($Q_{chain}$,$E$) space and not to their absolute values.

Thus, we find that a dramatic suppression of the INS magnetic intensity occurs in an antiferromagnetic cuprate as a result of strong covalent bonding between the Cu 3$d$ magnetic states and the 2$p$ states of the surrounding O ligands. We were able to isolate and unambiguously identify this effect thanks to the outstanding progress in experimental and theoretical techniques that has occurred over the past decade. The success of this study rested on combining (i) availability of large, high quality crystals, (ii) recent development of TOF neutron scattering instrumentation, (iii) exact Bethe ansatz results for S($Q$,$E$), and (iv) recent advances in the first principles calculations, including construction of the low-energy Wannier functions. The importance and future impact of our findings is hard to overemphasize. They provide important framework for understanding the existing INS results in cuprates and offer natural solution to the long-standing controversy of the missing INS intensity in cuprates[10,12,28]. Other positive conclusions are plenty and we spell out here but a few. Our results render direct experimental credibility to the *ab initio* form factor calculations in the cuprates and thus establish deployment of these new theoretical techniques as a useful tool for the INS analysis. They also emphasize the importance of the theoretical progress in the seemingly esoteric field of integrability-based calculations for 1D spin chains for solving very practical puzzle in the field of cuprate magnetism and superconductivity. Finally, perhaps the most spectacular result of this work is that INS is established as a sensitive experimental tool for probing not just spin correlations, but also important features of the electronic wave functions of unpaired magnetic electrons. Through the unique non-perturbing nature of the neutron probe and combined with the capabilities of new neutron sources currently being developed, this opens unprecedented new opportunities for the future.

**Methods**

The $Sr_2CuO_3$ sample used in our measurements was composed of three large single crystals with a combined mass of 18.45 g. All crystals were grown using the travelling solvent-floating zone method and had mosaic equal to or smaller than 0.3° full width at half maximum (FWHM). The crystals were mutually coaligned on an aluminium sample holder to better than 0.4° and 0.9° with respect to rotations around *a* and *c* lattice directions respectively. The sample was mounted on the cold head of the closed-cycle refrigerator in the evacuated scattering chamber of the MAPS spectrometer at the ISIS spallation neutron source at the Rutherford Appleton Laboratory, UK. The *b-c* plane was nearly horizontal and the *b* (chain) direction was aligned perpendicular to the incident neutron beam.

The INS spectrum of $Sr_2CuO_3$ was measured at four different incident neutron energies, 240 meV, 516 meV, 794 meV and 1088 meV and with the Fermi chopper spinning at 400 Hz, 400 Hz, 500 Hz and 600 Hz, respectively. Although the lower incident energies do not allow the full energy range of the spinon excitations to be measured, they provide better energy resolution on the lower energy excitations. Data were normalized by measuring the incoherent neutron cross-section of a vanadium sample of known mass in each configuration. For each cut, non-magnetic intensity arising from the incoherent and coherent multiple phonon and multi-phonon scattering by the sample and its environment was fitted to a simple smoothly varying functional form $A + B\Phi^2 + CQ_{chain}$ , where $\Phi$ is the angle by which neutrons are scattered, and that allows for a small empirically observed component linear in $Q_{chain}$. The cuts in Figure 3 show unsubtracted data; Figure 2 has had the fitted background subtracted. The fits of $S(\mathbf{Q}, E)$ to the data were performed using the TOBYFIT program (TGP, unpublished), which uses Monte Carlo integration to convolute the instrumental resolution function with models for the scattering cross-section.

The theoretical *ab initio* electronic structure calculation was performed with LDA+$U$ ($U = 8$ eV) approximation of density functional theory implemented in the WIEN2k code via full potential, all-electron, linearized augmented plane wave (LAPW) basis. In $Sr_2CuO_3$, the $d$ shell has a single hole in the $x^2$-$y^2$ orbital. The Wannier function is constructed from LDA+U orbitals, such that it preserves the local point group symmetry and spans the low-energy Hilbert space of the hole in the electronic structure[19,20]. This describes the magnetic electron density, correctly accounting for the hybridization with surrounding $p$ orbitals of the oxygen ligands. The magnetic form factor is the Fourier transform of the norm of this low-energy Wannier function.

**Acknowledgements** We acknowledge discussions with J. Tranquada, F. Essler, H. Benthien, and D. F. McMorrow. Work at BNL was supported by the U.S. Department of Energy. J.-S. C. acknowledges support from the FOM foundation.

**Competing Interests statement** The authors declare they have no competing financial interests.

**Correspondence** and requests for materials should be addressed to I. A. Z. (zaliznyak@bnl.gov) and A. C. W. (a.walters@ucl.ac.uk).

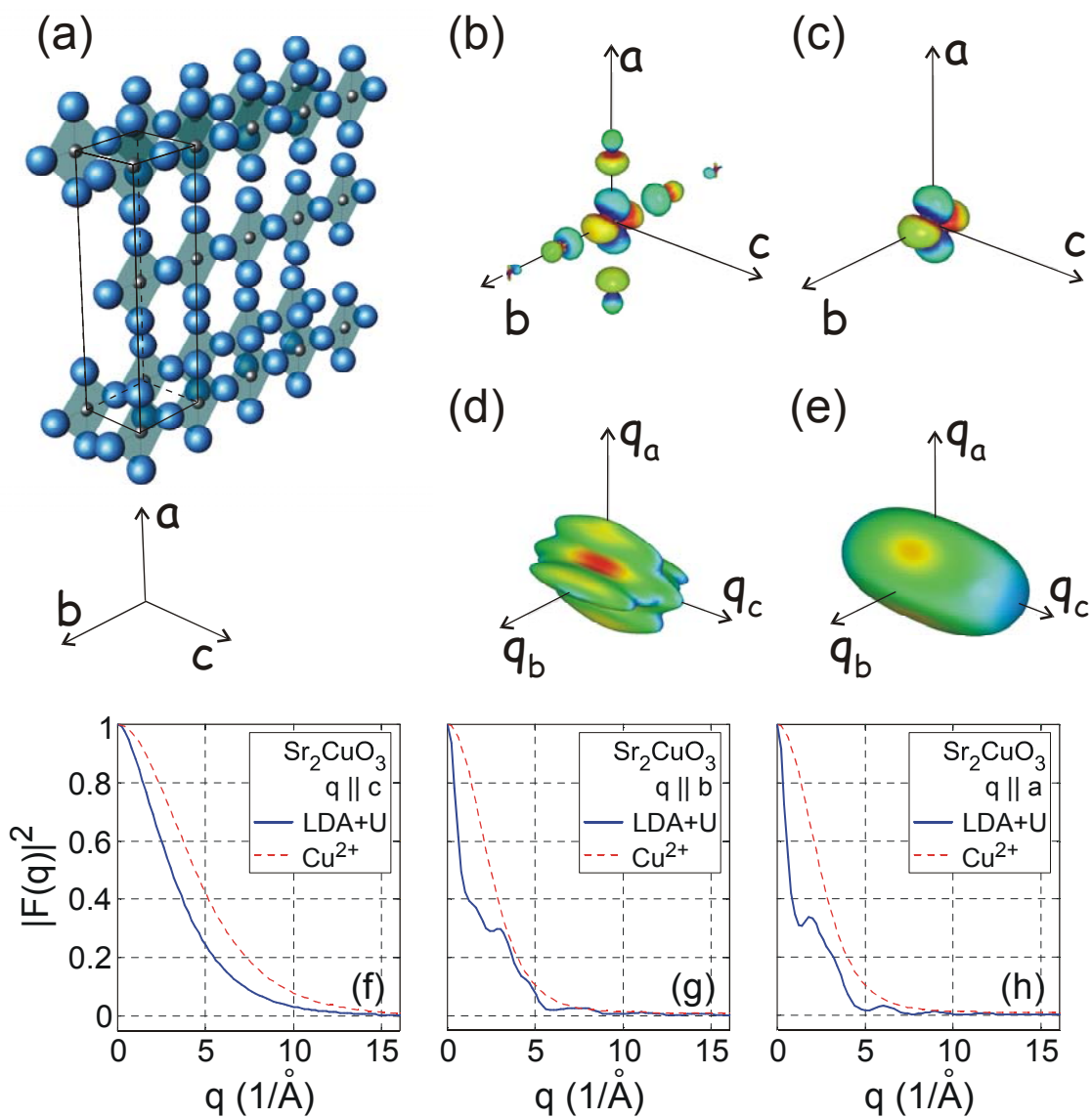

Figure 1 (a) Copper (small spheres) and oxygen (large spheres) ions in the
$Sr_2CuO_3$ crystal structure. Spheres show half of the corresponding ionic
radii. Corner-sharing $CuO_4$ square plaquettes form chains along the *b* axis.
(b) Covalently hybridized Wannier wavefunction of the unpaired magnetic
electron in $Sr_2CuO_3$ obtained from the *ab initio* LDA+U calculation. The
equal-density surface at $|\psi(\mathbf{r})|^2$=0.05 Å$^{-3}$ is shown. (c) Similar depiction for
the $Cu^{2+}$ ionic wave function of the $3d(x^2-y^2)$ orbital typically used for magnetic form factor calculations[11,12,25,26]. (d) and (e) show the equal-level surfaces of the magnetic form factor squared at $|F(\mathbf{q})|^2 = 0.13 \approx 1/e^2$ for the wave functions in (a) and (b), respectively. (f)-(g) compare cuts of magnetic form factors shown in (d) [solid blue line] and (e) [broken red line] along the three principal crystal axes.

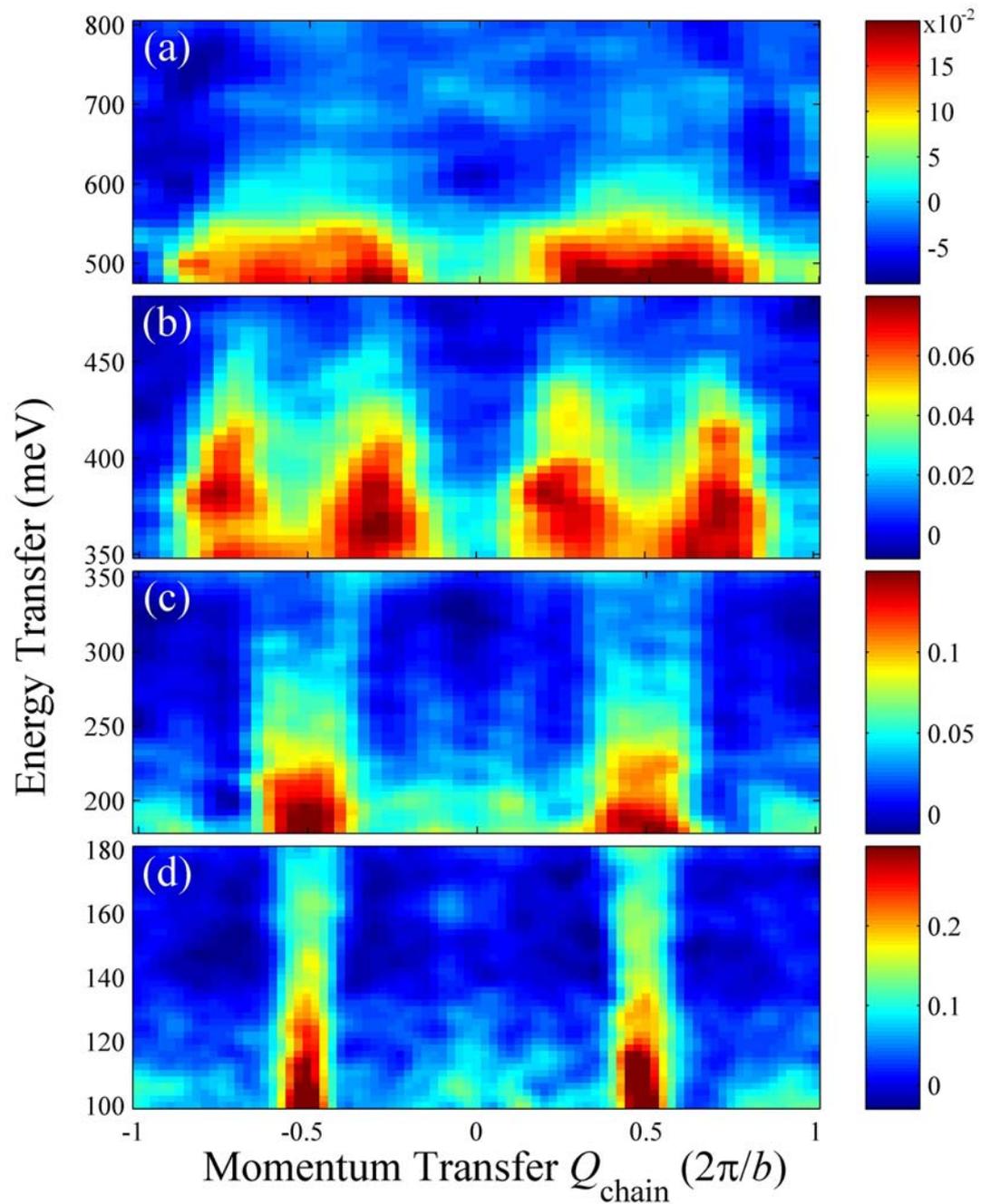

Figure 2 Inelastic neutron scattering intensity from $Sr_2CuO_3$ plotted as a function of momentum parallel to the chains and energy transfer. In direct geometry TOF neutron spectrometers such as MAPS, a pulse of monochromatic neutrons with energy $E_i$ is incident on the sample and then

the energy lost or gained by the neutron in the sample is ascertained by measuring the time taken for the neutrons to travel to the detector. By using time-resolved detectors, the whole energy transfer range E < $E_i$ is measured in a single pulse. Optimization of resolution and intensity conditions required using four different incident neutron energies $E_i$ shown in panels (a) 1088 meV, (b) 794 meV, (c) 516 meV and (d) 240 meV. A choice of lower $E_i$ results in superior momentum and energy resolution, but can only measure excitations up to $E_i$. The data for each $E_i$ were accumulated for one to several days. The intensity is in mbarn/Steradian/meV/Cu ion and color scale in each panel is adjusted to emphasize magnetic scattering.

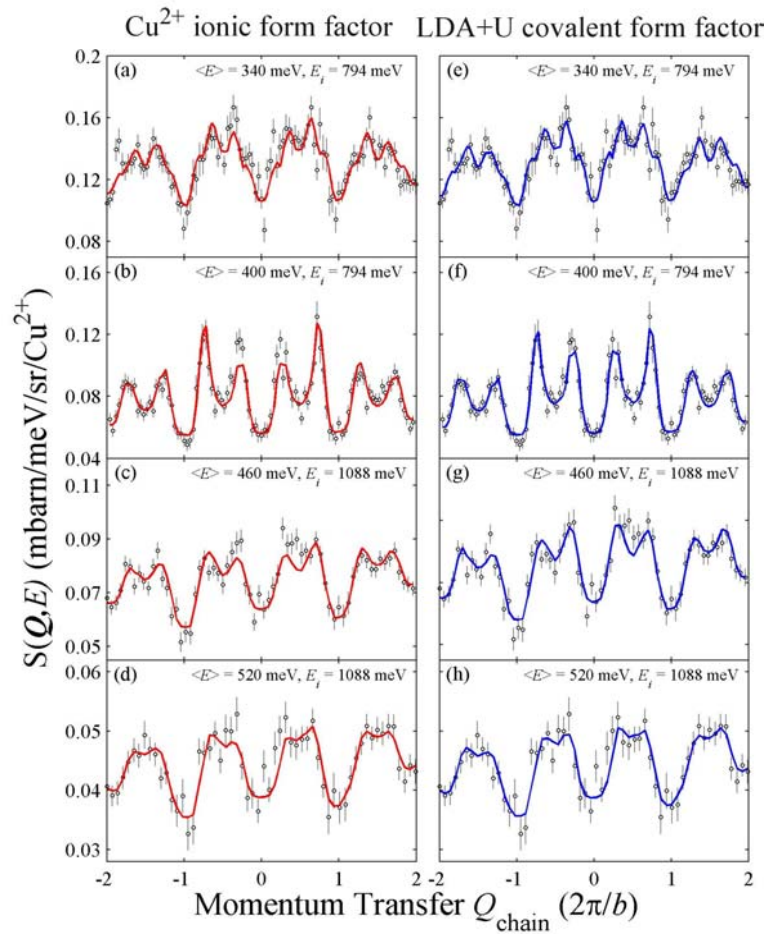

Figure 3 Selected constant energy cuts through the data as a function of momentum transfer parallel to Cu-O chains. Lines are fits to the exact two- and four-spinon scattering function for the spin-½ Heisenberg chain, using the ionic $Cu^{2+}$ magnetic form factor of Figure 1 (d) (left column) and the LDA+U covalent magnetic form factor of Figure 1 (c) (right column). The in-chain exchange coupling and the intensity prefactor were the only parameters varied in these fits. Similar fits were also performed for many additional constant energy cuts. The better agreement achieved by using the covalent form factor is visible when comparing panels (f) with (b), (g) with (c) and (h) with (d).

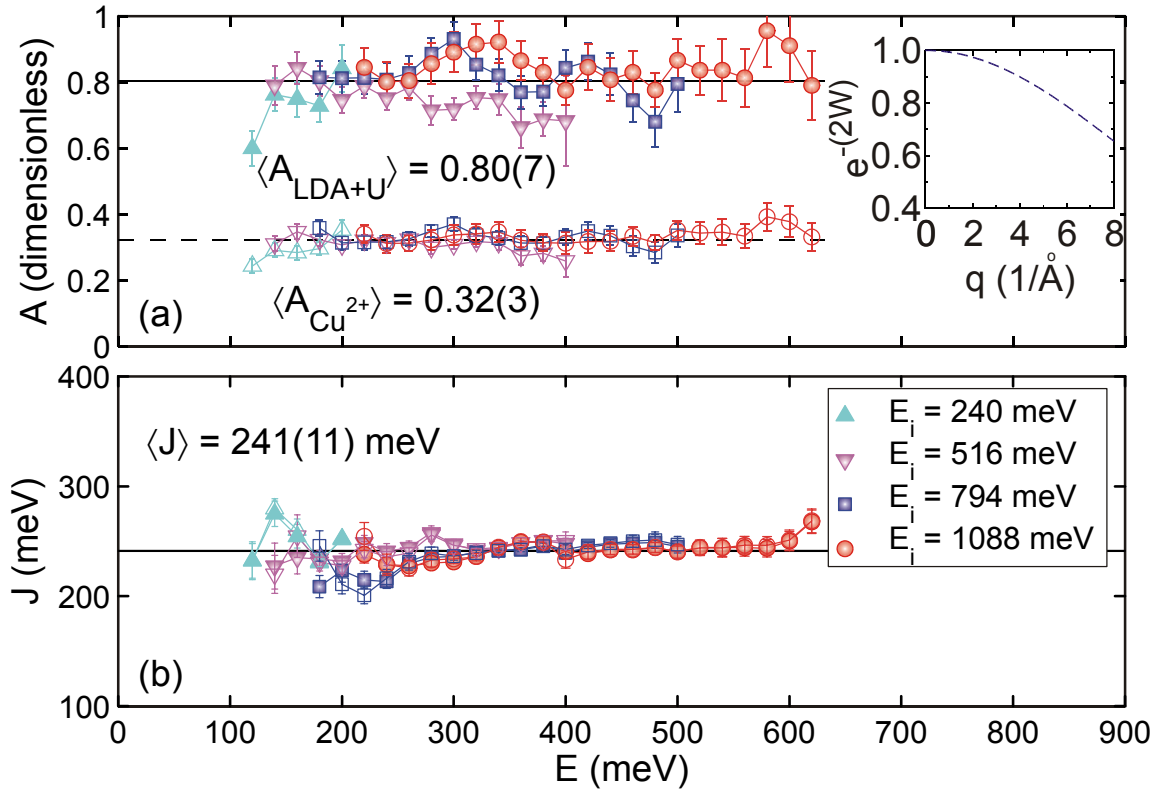

Figure 4 (a) The intensity prefactor A and (b) the exchange integral J, obtained by fitting a number of constant-energy scans collected with different incident energies, such as shown in Figure 3, to the exact expression for the sum of 2- and 4-spinon scattering and using the LDA+U covalent magnetic form factor (filled symbols) and that of the free $Cu^{2+}$ ion (open symbols). Inset in panel (a) illustrates the magnitude of additional intensity suppression by the Debye-Waller factor, calculated for bulk copper[16].